\documentclass[12pt]{article}           

\def\preprint{HUB-EP-00/14}       
\def\finished{February 2000}
\def\archive {hep-th/0002246}           

\def\title{Reflexive Polyhedra and their \\[5mm]
           Applications in String and F-theory}

\long\def\abstract{This is an informal introduction to the concept of reflexive
 polyhedra and some of their most important applications in perturbative and
 non-perturbative string physics. Following the historical
 development, topics like mirror symmetry, gauged linear sigma models,
 and the geometrical structures relevant to string and F-theory dualities are
 discussed. Finally some recent developments concerning the classification of
 reflexive polyhedra are mentioned.
 }

\input epsf

\usepackage{amssymb,latexsym}              \catcode`\"=\active \let"=\"

\def\ifundefined#1{\expandafter\ifx\csname#1\endcsname\relax}
\def\bye{\end{document}}   
\long\def\new#1\endnew{{\bf #1}}

\def\HS#1 {\hspace*{#1pt}} \def\VS#1 {\vspace*{#1pt}} \long\def\del#1\enddel{} 
\def\BC{\begin{center}}    
\def\EC{\end{center}}

\let\0=\over      \let\1=\vec      \def\2{{1\over2}}    \let\3=\ss
\let\4=\underline \let\5=\overline \let\6=\partial      \def\7#1{{#1}\llap{/}}
\def\8#1{{\textstyle{#1}}}         \def\9#1{{\ifmmode{\pmb{#1}}\else\bf#1\fi}}

          \def\({\left(}       \def\){\right)}

\def\eeql#1 {\label{#1}\eeq}        
\def\beq{\begin{equation}}      \def\eeq{\end{equation}}        
\def\bea{\begin{eqnarray}}      \def\eea{\end{eqnarray}}

\let\and=\wedge                   
      \let\iff=\Leftrightarrow
                
\let\bra=\langle        \let\ket=\rangle        \def\<#1\>{\bra #1 \ket}

\def\rel#1 #2{\buildrel #1 \over {#2}}  
\def\fnote#1#2{\begingroup\def\thefootnote{#1}\footnote{#2}
                \addtocounter{footnote}{-1}\endgroup}   

\def\subdef#1{\gdef\globalColor##1{##1}}      
      \subdef{Black}

   \let\l=\lambda

             \let\D=\Delta

\def\IR{{\mathbb R}} \def\IC{{\mathbb C}} \def\IP{{\mathbb P}} 
   
\def\IZ{{\mathbb Z}}

\def\plb#1 #2 {Phys. Lett. {\bf B#1} #2 }
\def\phr#1 #2 {Phys. Rep. {\bf  #1} #2 }        
\def\npb#1 #2 {Nucl. Phys. {\bf B#1} #2 }
\def\aph#1 #2 {Ann. Phys. {\bf #1} #2 }         
\def\jmp#1 #2 {J. Math. Phys. {\bf #1} #2 }
\def\jgp#1 #2 {J. Geom. Phys. {\bf #1} #2 }
\def\prd#1 #2 {Phys. Rev. {\bf D#1} #2 }
\def\prl#1 #2 {Phys. Rev. Lett. {\bf #1} #2 }
\def\rmp#1 #2 {Rev. Mod. Phys.  {\bf #1} #2 }
\def\zpc#1 {Z. Phys. {\bf #1C} }
\def\cmp#1 #2 {Commun. Math. Phys. {\bf #1} #2 }
\def\cqg#1 #2 {Class.Quant.Grav. {\bf #1} #2 }
\def\mpl#1 {Mod. Phys. Lett. {\bf A#1} }
\def\cpc#1 {Computer Phys. Commun. {\bf #1} }   % Belfast,cpc@v1.am.qub.ac.uk
\def\ijmp#1 {Int. J. Mod. Phys. {\bf A#1} }
\def\ijmpC#1 {Int. J. Mod. Phys. {\bf C#1} }

\def\BP{\begin{picture}} \def\EP{\end{picture}}         %% --> PICTURE macros 
\newcounter{TRefNX} \let\OLDcite=\cite  \makeatletter%       DRAFT MODE macros
\def\makeTRefs#1{\@for  \NewTRef:=#1\do{\global\makeTRef{\NewTRef}}}
\def\makeTRef#1{\ifundefined{TRef#1}\stepcounter{TRefNX}%
\expandafter\xdef\csname TRef#1\endcsname{\theTRefNX}\fi}\makeatother
\def\NEWcite#1{\makeTRefs{#1}\OLDcite{#1}}  

\ifundefined{draftmode} {} \else        \let\cite=\NEWcite
\newcount\HOUR  \HOUR=\the\time \divide\HOUR by 60      \multiply \HOUR by 60 
\newcount\MIN   \MIN=\the\time  \advance\MIN by -\HOUR  \divide\HOUR by 60
\ifnum  \MIN>9  \def\printTIME{{\it\the\HOUR\,:\,\the\MIN}}
        \else   \def\printTIME{{\it\the\HOUR\,:\,0\the\MIN}} \fi % \printTIME

\def\LLab#1{\BP(0,0)\unitlength=1mm\put(-12,.5){\makebox(0,0)[cr]{\small #1
        \rlap{$_{_{\makeatletter\csname TRef#1\endcsname\makeatother}}$}}}\EP}
\fi

\begin{document}

{\hfill \archive   \vskip -2pt \hfill\preprint }
\vskip 15mm
\begin{center}  {\huge\bf   \title }\end{center}
\begin{center} \vskip 10mm
Harald SKARKE\fnote{\#}{e-mail: skarke@physik.hu-berlin.de}
\\[3mm] Institut f\"ur Physik, Humboldt-Universit\"at zu Berlin\\
        Invalidenstra\ss e 110, D-10115 Berlin, Germany
        
\vfill                  {\bf ABSTRACT } \end{center}    \abstract

\vfill \noindent \preprint\\[5pt] \finished \vspace*{9mm}
\thispagestyle{empty} 

\newpage

\pagestyle{plain}
\setcounter{page}{1}
  
%\enddel

\section{Perturbative string theory and reflexive polyhedra}

Originally the introduction of reflexive polyhedra was motivated by mirror
symmetry:
A superstring compactification whose target space is a Calabi--Yau threefold  
leads, on the string world sheet, to an $N=2$ superconformal field theory with
central charge $c=9$.
The $N=2$ superalgebra of this world sheet theory decomposes into a right and
a left moving part, each of which has a $U(1)$ R-symmetry.
A change in the relative signs of the left/right $U(1)$ charges is a trivial
operation from the point of view of world sheet dynamics, but it radically
changes the space--time interpretation of the model:
Assuming that the model has a geometrical interpretation before and
after the sign flip, this implies that we have passed from a Calabi--Yau
manifold to a different Calabi--Yau manifold in such a way that the Hodge
numbers $h_{11}$ and $h_{12}$ are interchanged.
The conjecture that most Calabi--Yau manifolds have mirrors of this kind was
supported by the first explicit constructions of large classes of such
manifolds as hypersurfaces in weighted projective spaces \cite{CLS}: 
It was found 
that for most of the resulting Hodge pairs the mirror pair could also be found.
The completion of the classification of Calabi--Yau hypersurfaces in weighted 
projective spaces \cite{KlS,nms} showed, however, that not all of the mirrors 
could be found within the same class.
The resolution to this problem was provided by the introduction of 
Calabi--Yau manifolds that are constructed as hypersurfaces in
toric varieties with the help of so--called reflexive polyhedra \cite{bat}.
This generalized framework provided not only the missing mirrors but also
many completely new models.

Reflexive polyhedra are defined with respect to a dual pair of lattices 
$M\simeq \IZ^n$ and $N\simeq\IZ^n$ and the underlying real vector spaces 
$M_\IR\simeq \IR^n$ and $N_\IR\simeq\IR^n$.
They are polytopes in these real vector spaces with the origin in their
respective interiors.
For any such polytope $\Delta\subset M_\IR$ one can define the dual polytope as
\beq \Delta^*:=\{v\in N_\IR:\;\<v,w\>\ge -1 \;\forall\;w\in \D\}. \eeq
A lattice polyhedron $\Delta$ is a polyhedron in $M_\IR$ with vertices in $M$,
and a reflexive polyhedron is a lattice polyhedron $\D$ with {\bf 0} in its 
interior such that $\D^*$ is also a lattice polyhedron.

{}From such polytopes, complex manifolds can be constructed in the following 
simple way:
One draws rays $v_i$ through the vertices (or, more generally, through 
lattice points) of $\D^*$ and introduces a homogeneous coordinate $x_i$ for 
every ray $v_i$ in a way similar to the construction of projective space.
Then one must find a complete set of linear relations of the type 
$\sum q_iv_i=0$ among the lattice vectors defining these rays, and for every 
such linear relation one
introduces a multiplicative equivalence relation $x_i\simeq \l^{q_i}x_i$ 
among the homogeneous coordinates.
\begin{figure}[htb]
\epsfxsize=2.5in
\hfil\epsfbox{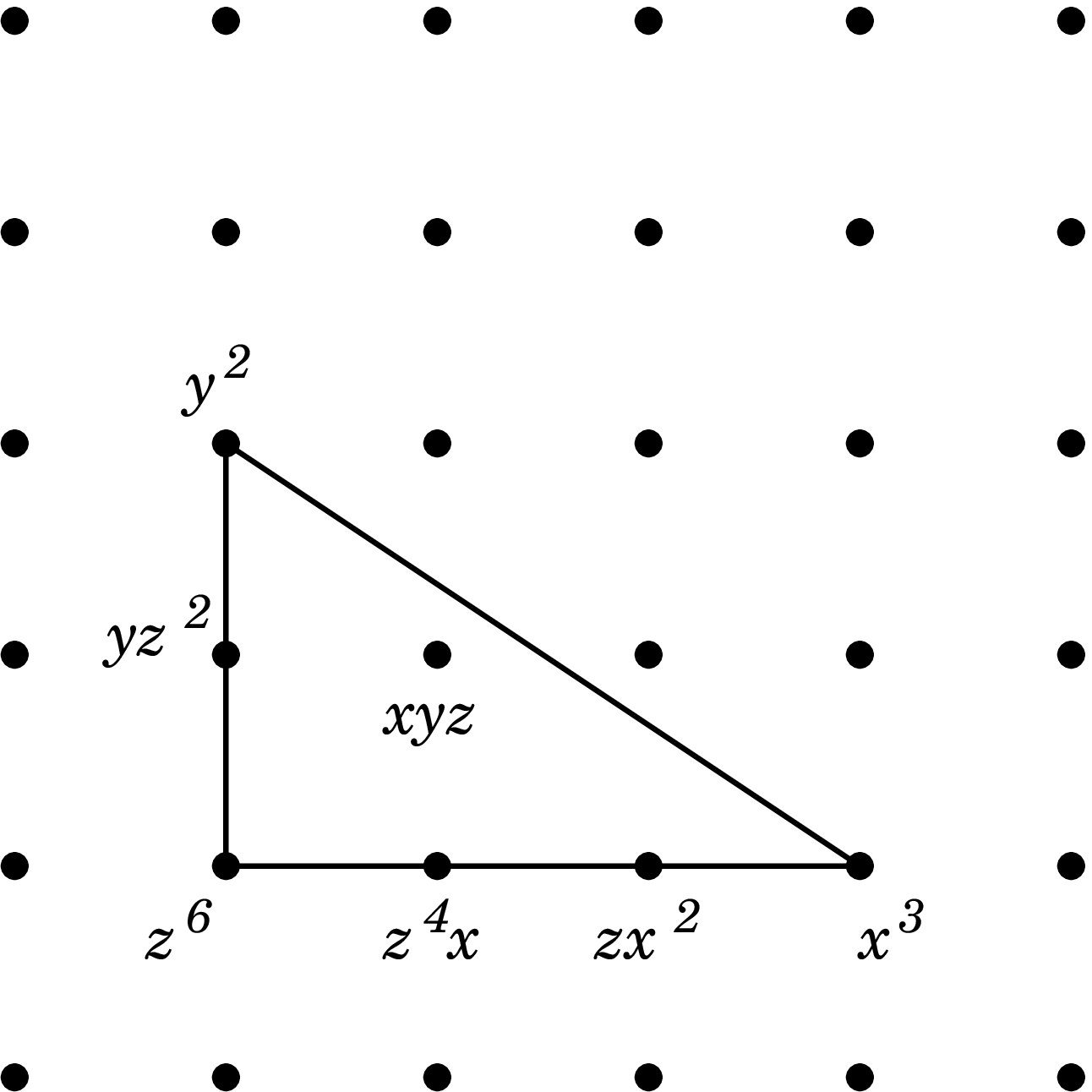}\hfil
\epsfxsize=2.5in
\hfil\epsfbox{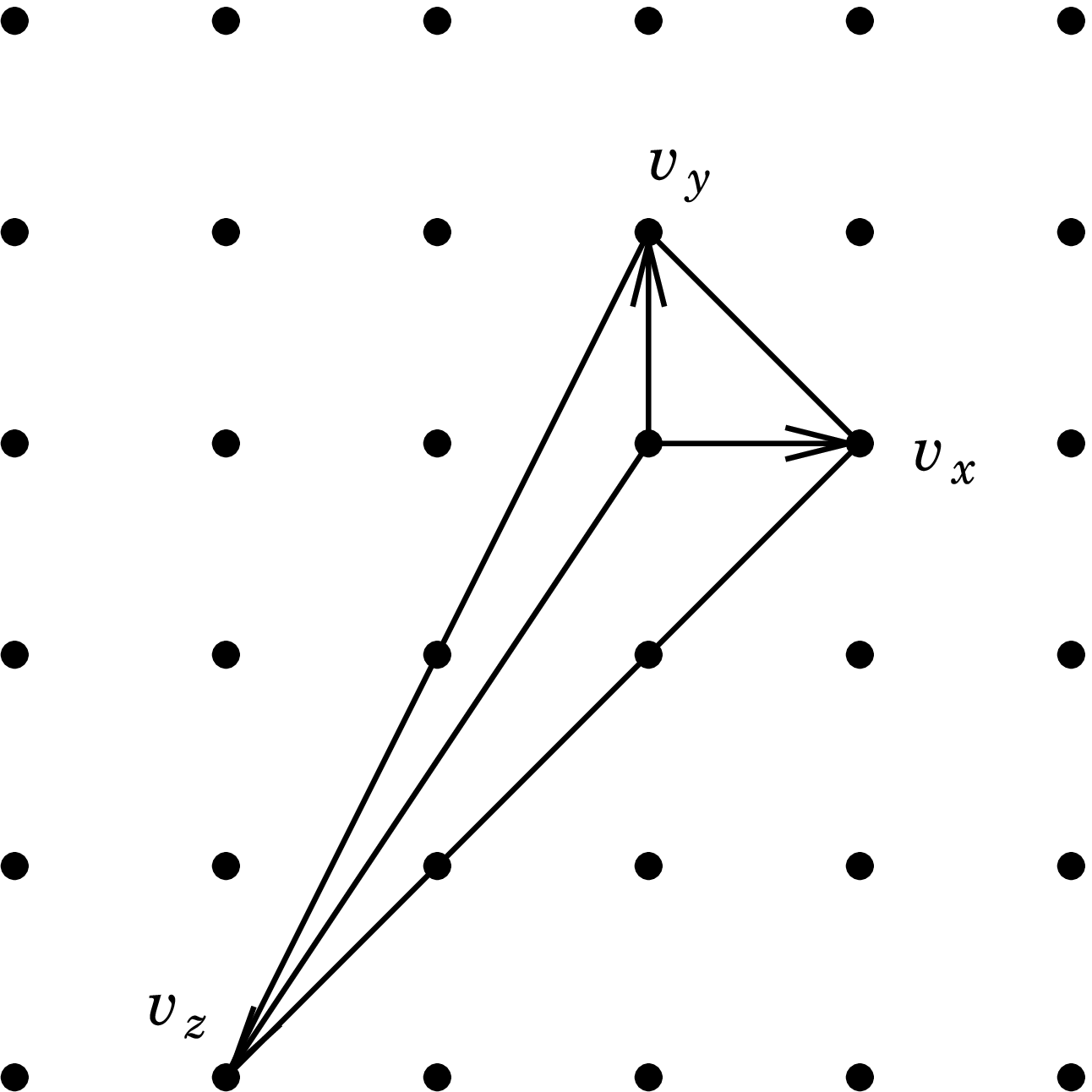}\hfil
\caption{$\D$ and $\D^*$ for a weighted projective space}
\label{fig:fans}
\end{figure}
This is most easily visualized with an example: For the reflexive pair
of fig. 1, the rays corresponding to the vertices of $\D^*$ fulfil
$2 v_x+ 3 v_y+ v_z=0$, so the space described in this way is just
the weighted projective space defined as the set of equivalence classes in
$\IC^3\setminus\{0\}$ with respect to the relations 
$(x,y,z)\simeq (\l^2x, \l^3y, \l z)$.
It is not hard to see that this space suffers from singularities at the points
$y=z=0$ and $x=z=0$.
In algebraic geometry there is a procedure for turning such a singular space
into a smooth one, referred to as `blowing up' singularities.
In the toric setup, it is very easy to perform this operation: It simply
corresponds to adding extra rays to the toric diagram; in the present example
these are just the rays through the lattice points interior to edges of
$\D^*$.

In order to define hypersurfaces in toric varieties, it is necessary to have
equations that are compatible with the equivalence relations among the toric
coordinates.
For the case of Calabi--Yau hypersurfaces, these homogeneous equations are
particularly simple.
In terms of rays $v_i$, corresponding coordinates $x_i$ and lattice points $w$
of $\D$, a Calabi--Yau hypersurface is determined by
\beq P(x_1,\ldots,x_k)=
     \sum_{w\in\D\cap M}a_w \prod_{i=1}^k x_i^{\<v_i,w\>+1}=0,  \eeq
i.e. every lattice point of $\D$ determines a monomial in $P$.
For our example, which leads to a one dimensional Calabi--Yau hypersurface
which is nothing but an elliptic curve, the corresponding monomials are 
indicated in fig.1.
In the context of polyhedra mirror symmetry manifests itself as the exchange 
of $(M, \D)$ and $(N,\D^*)$.

Almost immediately after its introduction this construction turned out to be
important for further applications in the context of string theory:
Witten's gauged linear sigma models \cite{Wph} are $d=2$ theories with $N=2$
supersymmetry.
In such a model a chiral superfield is introduced for every toric coordinate 
and $U(1)$ gauge fields implement the compact parts of the multiplicative 
relations.
The superpotential is just the previously introduced homogeneous
polynomial multiplied with an auxiliary field.
At the fixed points of the renormalization group flow, such a model is
believed to possess superconformal symmetry.
Within a certain range of the coupling constants the following
scenario takes place:
The equations of motion constrain the superfields in such a way that the
noncompact part of the relations is also fixed and the homogeneous
polynomial is forced to vanish, leading effectively to a Calabi--Yau sigma
model.
For other values, one gets a Landau--Ginzburg model or some hybrid.
As shown by Aspinwall, Greene and Morrison \cite{AGM1,AGM2} this construction 
implies that the corresponding string physics allows for physically smooth 
topology changing transitions.

\section{String dualities and reflexive polyhedra}

One of the main new insights of the second string revolution is that there
appear to be non-perturbative dualities between seemingly different theories.
Examples of this type are the dualities between heterotic string theory
compactified on $K3\times T^2$ and type IIA on a $K3$ fibered 
Calabi--Yau threefold
and between heterotic string theory compactified on Calabi--Yau $n$-folds and 
F-theory on elliptically fibered Calabi--Yau $(n+1)$-folds.
While heterotic string theory has non-abelian gauge groups in its perturbative
spectrum, the dual theories can develop them only if the compactification
manifold becomes singular. 
Certain singularities allow for a classification that follows exactly the
pattern of the classification of simply laced Lie groups, and the
non-perturbatively enhanced gauge groups are just the corresponding
ADE gauge groups.
Thus the two most important geometrical ingredients for making these dualities
work are fibration structures and singularities.
As we will now see, both types of structures manifest themselves in simple
ways in the context of reflexive polyhedra.

A manifold has a fibration structures if there is a projection to some other
manifold called the base manifold, and if the preimage of a generic point of
the base under this projection is isomorphic to another manifold which is the
fiber. 
In terms of reflexive polytopes, the fiber corresponds to a reflexive
subpolytope $\D^*_{\rm fiber}\subset \D^*_{\rm total}$ in a linear subspace 
$N_{\IR, \rm fiber}\subset N_{\IR}$ containing {\bf 0}.
The base space is just the toric variety whose fan (i.e., collection of rays 
and some other data) is the projection of the complete fan along the subspace
$N_{\IR, \rm fiber}$.
\begin{figure}[htb]
\epsfxsize=3in
\hfil\epsfbox{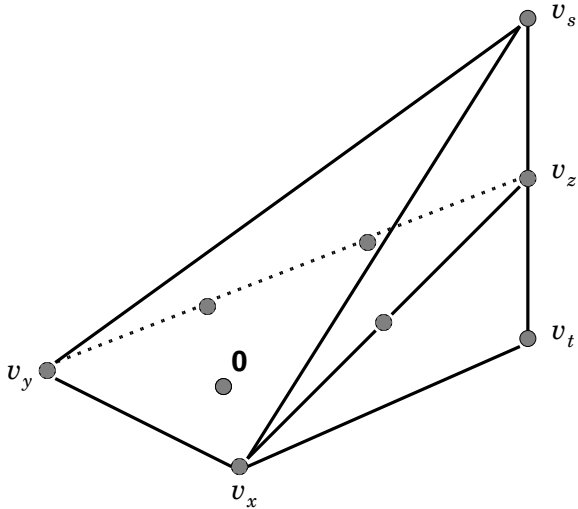}\hfil
\caption{The polyhedron $\D^*$ corresponding to a smooth elliptically fibered
K3 surface}
\label{fig:wei}
\end{figure}
As an example, fig. \ref{fig:wei} shows the fan of a smooth elliptically 
fibered K3 surface, with the fiber determined by the triangle $v_xv_yv_z$.
The toric diagram for the base space $\IP^1$ is determined by projecting along
the plane of the triangle. It is a line segment isomorphic to $v_sv_t$.

\begin{figure}[htb]
\epsfxsize=2.5in
\hfil\epsfbox{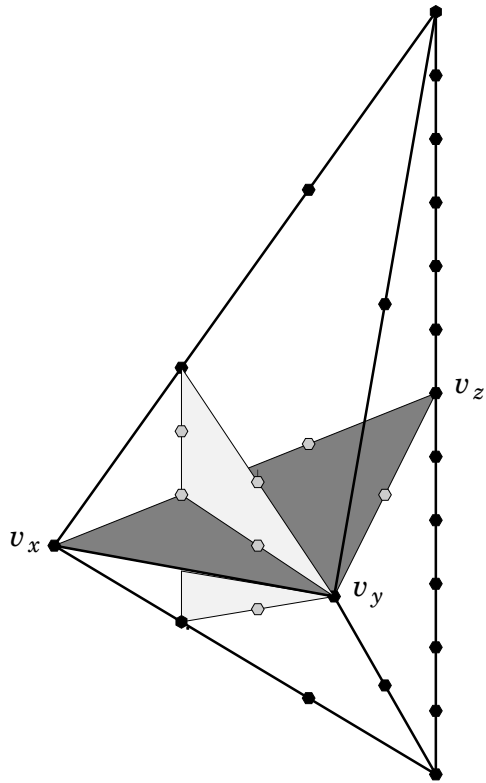}\hfil
\caption{The polyhedron $\D^*$ corresponding to the heterotic string with 
unbroken gauge symmetry}
\label{fig:poly}
\end{figure}

ADE singularities manifest themselves in a particularly elegant way in
terms of reflexive polyhedra: 
As Candelas and Font observed \cite{CF}, under favourable
circumstances the Dynkin diagrams of non-perturbative gauge groups can be 
seen in the polyhedra corresponding to the blow-up of the singular space.
As an example fig. \ref{fig:poly} shows the reflexive polytope
that provides the F-theory dual of heterotic string theory compactified to
eight dimensions with an unbroken gauge group of $E_8\times E_8$.
The horizontal triangle represents the elliptic fiber, and the diagram
obtained by considering the lattice points of $\D^*$ above it and the edges 
connecting them is nothing but the extended Dynkin diagram of $E_8$ 
(the other $E_8$ is visible in the same way below the triangle).
Using this polytope for constructing threefolds with the corresponding K3
manifold as a fiber, it was possible to obtain theories in six dimensions 
with gauge groups like 
$E_8\times(E_8\times F_4\times G_2^2\times A_1^2)^{16}$ \cite{CPR1}.
This was actually the `record gauge group' until Aspinwall and Morrison
constructed a model with the group
$SO(32)\times Sp(24)\times SO(80)\times Sp(48)\times SO(128)\times Sp(72)
   \times SO(176)\times Sp(40)\times Sp(52)\times SO(64)$, using $SO(32)$
heterotic string theory.
This presented the challenge of finding the F-theory dual of the $SO(32)$ 
heterotic string.
Remarkably, it is represented again by the polytope of fig. \ref{fig:poly}
\cite{fst}.
This time the fibration structure is determined by the vertical triangle and
the lattice points to the right of it are connected by the edges into the
extended Dynkin diagram of $SO(32)$.

\section{Classification of reflexive polyhedra}

Given all these interesting properties of reflexive polyhedra, one may
certainly wonder how many of them exist in a given dimension $d$.
Until 1998, this was known only for the case of $d=2$ with 16 reflexive 
polygons.
This situation has changed with the development of a classification scheme that
works (in principle) in arbitrary dimensions \cite{crp,wtc}.
The main ideas of this scheme are as follows: 
Clearly it is enough to find a finite set of polyhedra that contain all 
reflexive polytopes as subpolytopes.
By duality, $\D_1\subset\D_2 \iff \D_2^*\subset\D_1^*$, so this task is
equivalent to finding a set of polyhedra such that every reflexive polytope
contains one of them.
In this spirit we define a minimal polyhedron $\nabla\subset N_\IR$ as a 
polyhedron with
{\bf 0} in its interior such that the convex hull of any proper subset of 
vertices of $\nabla$ fails to have {\bf 0} in its interior.
It can be shown that the vertices of minimal polyhedra always belong to 
(possibly lower dimensional) simplices with {\bf 0} in their respective 
interiors, and the different simplex structures can be classified.
In the same way as in the first section, these simplices determine weight 
systems corresponding to the linear relations of the vertices.
The weight systems of the simplices in a minimal polytope
together form a combined weight system which uniquely specifies the linear
structure of $\nabla$, but not the lattice.
The additional restrictions that $\nabla$ should be a lattice polytope and 
that the convex hull of the intersection of $\nabla^*$ with $M$
should have {\bf 0} in its interior lead to finite numbers of combined weight 
systems.
Thus all reflexive polyhedra can be obtained as subpolyhedra of the convex
hulls of $\nabla^*\cap M$, where $\nabla$ is given by some combined weight
system.

There are certain subtleties in this approach: One should take into account
that the lattice is not always uniquely determined by the weights, but there 
are methods to deal with sublattices of the finest possible $M$
lattice which is dual to the lattice generated by the vertices of $\nabla$.
Given the fact that two polytopes should be considered equivalent if
they are related by lattice isomorphisms it is necessary to define normal forms
of polyhedra that take this difficulty into account. 
Finally, in the case of four dimensional reflexive polyhedra 
%leading to Calabi--Yau threefolds, 
one nearly encounters the limits of today's available 
computer power: Unless numerous tricks are used, there will be numerical
overflows, lack of RAM and disk space, and an enormous computation time.

Nevertheless, the algorithm works and has produced the following results:
In $d=3$, there are 4319 reflexive polyhedra corresponding to K3 surfaces
\cite{c3d} which can be produced by a modern PC within 8 seconds.
For the case of $d=4$, the programs were still running at the time this talk
was given and had produced around 300 million polyhedra.
At that time I estimated the total number to be around $0.5\pm 0.1$ GCY
(Giga--Calabi--Yau).
This estimate was confirmed by the recent accomplishment of this project
after a computation time of more than half a year on several processors:
There are 473,800,776 reflexive polyhedra giving rise to
30,108 distinct pairs of Hodge numbers for Calabi--Yau threefolds \cite{c4d}.

{\it Acknowledgement:} 
The European Union TMR project ERBFMRX-CT-96-0045 is not only the organizer of
the conference where this talk was delivered, but has also supported my 
research since November 1998. Thanks!

\bye